\documentclass[aps,showpacs,prc,floatfix,twocolumn]{revtex4}

\usepackage{graphicx}
\usepackage{amssymb}

\newcommand{\en}{{\mathcal N}}
\newcommand{\zet}{{\mathcal Z}}
\newcommand{\dii}{{\rm d}}
\newcommand{\Tr}{{\rm Tr}}
\newcommand{\ii}{{\rm i}}
\newcommand{\bsq}{(B,S,Q)}

\begin{document}

\title{Chemical factors in canonical statistical models for
        relativistic heavy ion collisions}
\author{A. Ker\"anen}
\email{antti.keranen@oulu.fi}
\affiliation{Department of Physical Sciences,\\
P.O. Box 3000, FIN-90014 University of Oulu, Finland}
\author{F. Becattini}
\email{becattini@fi.infn.it}
\affiliation{Universit\`a di Firenze and INFN Sezione di Firenze,\\
Via G. Sansone 1, I-50019 Sesto F.no, Firenze, Italy}


\begin{abstract}
We study the effect of enforcing exact conservation of charges in statistical 
models of particle production for systems as large as those relevant to 
relativistic heavy ion collisions. By using a numerical method developed
for small systems, we have been able to approach the large volume limit 
keeping the exact canonical treatment of all relevant charges, namely baryon 
number, strangeness and electric charge. Hence, we hereby give the information 
needed in a hadron gas model whether the canonical treatment is necessary or 
not in actual cases. Comparison between calculations and experimental particle
multiplicities is shown. Also, a discussion on relative strangeness chemical
equilibrium is given.
\end{abstract}
\maketitle

\section{Introduction}\label{intro}

Over the era of relativistic nuclear collisions, the statistical-thermal models 
have been widely used in the analysis of particle production. During the past
few 
years, these models have been successful in describing particle multiplicities 
in high energy nuclear reactions at the stage where inelastic collisions between
hadrons cease (chemical freeze-out), see e.g. \cite{features, cleymans319,
heppe, 
cleymans5284, cleymans3319}.

The easiest to handle, thence the most used, statistical model calculational
framework is based on the grand canonical (GC) ensemble. In the GC approach
entropy is maximized using the constraints of conserved {\em ensemble averages}
of
energy and charges. This allows the net charges to fluctuate from sample to 
sample even though the actual charges brought into the physical reactions
were exactly same every time. In the large volume and energy limit, the error 
in such an assumption due to charge fluctuations is negligible.

On the other hand, when the involved volumes are small, like in e$^+$e$^-$, 
p$\bar{\mbox{p}}$, pA reactions or in peripheral AA reactions, the charge
fluctuations 
allowed in GC formalism give a large theoretical error. So, the GC ensemble
turns out to be inadequate for the analysis of experimental results and the 
statistical ensemble to be used is rather the canonical one, enforcing exact 
conservation of relevant charges. In this case entropy is maximized using the 
constraint of conserved ensemble average of energy, but charges are not allowed 
to fluctuate.

Statistical models based on the canonical ensemble have been able to reproduce
particle multiplicities even in elementary collisions, although in those there 
is clearly no room for kinetic thermalization \cite{becattini485,becattini269}.
Also, the strangeness enhancement with increasing system size in p--A reactions
can be explained by a canonical effect \cite{cleymans2747}.

In this work, we investigate whether and to what extent the canonical treatment 
is relevant for actual heavy ion collisions. The main difficulty of canonical
calculations is related to the involved large values of baryon number and
electric charge, which make the numerical computation of canonical partition 
functions quite hard. We present here an efficient numerical method to carry 
out such calculations which enabled us to implement exact charge conservation up
to
baryon 
number ${\cal O}(100)$ whereas with old methods it was only possible to reach
$B\sim 20$ \cite{cleymans2747} with very long computation times.

After deriving the needed expressions of mean particle numbers, we present a 
systematical study of the chemical factors appearing in the canonical
statistical
model. Along with some comparison between grand-canonical and canonical
calculations, 
we show a comparison with experimental results for peripheral to central A--A 
reactions at AGS and SPS energies.

In section \ref{gamma}, we briefly address the role of the so-called relative 
strangeness chemical equilibrium, usually parametrized with  $\gamma_S$, in
canonical 
statistical-thermal models.

\section{The analytical development}\label{model}


In nuclear physics, the problem of calculating the relativistic
canonical partition function has been handled by several different
methods, see e.g. \cite{bohr-mott,jennings440,chase2339,dasgupta1361}.
Here, we employ a very general group theoretical method 
-- applicable to any internal
symmetry represented by a semi-simple Lie algebra --
first introduced by Cerulus \cite{cerulus, turko201, turko153}.


By denoting the set of conserved quantum numbers by $\{C_i\}$, the canonical
partition function $Z_{\{C_i\}}$ can be obtained
from the usual grand-canonical partition function $Z_{GC}$
 by using a projecting operator 
onto the conserved quantum numbers. In the case of $N$ internal symmetries of
type
$U(1)$, the projection takes the form:
\begin{equation}
Z_{\{C_i\}}(T,V) = \left[ \prod_{i=1}^{N} 
\frac{1}{2\pi}  \int_0^{2\pi}\dii\phi_i e^{-\ii C_i\phi_i}\right]
         Z_{GC}(T,V,\{\lambda_{C_i}\}).
\label{eq:projection}
\end{equation}
where $\phi_i \in [0,2\pi)$ is a $U(1)$ group parameter and a Wick-rotated 
fugacity factor $\lambda_{C_i} = e^{i\phi_i}$ is introduced, for every charge 
$C_i$. 

In order to calculate $Z_{GC}$ in the above integral, we will assume the
Boltzmann statistics for all hadron species. The error involved in such an 
approximation
for final particle multiplicities, when resonance decays is taken into account, 
is some percent in case of pions and much less for all other hadrons. 
Quantum statistics may be used, but requires somewhat heavier calculations and
disregarding it does not essentially affect any forthcoming argument.

In heavy ion reactions, the relevant set of conserved charges is $\{C_i\} = B,
S, Q$, namely baryon number, strangeness and electric charge. Denoting the
partition
function of hadrons carrying none of the previous charges by $Z_0$, and writing 
$Z_{GC}$ by using the one-particle partition function $z_i^1 =
[(2J_i+1)V/(2\pi)^3] \int \dii^3 p \exp [-\sqrt{p^2+m_i^2}/T]$ for each hadron 
$i$, we find 
\begin{widetext}
\begin{eqnarray}
Z_{B,S,Q}(T,V) &=&  
Z_0 \frac{1}{(2\pi)^3}  \int_0^{2\pi}\dii\phi_B \; e^{-\ii B\phi_B}
\int_0^{2\pi}\dii\phi_S \; e^{-\ii S\phi_S}
\int_0^{2\pi}\dii\phi_Q \; e^{-\ii Q\phi_Q} \nonumber \\
&\times& \exp\left\{\sum_i z_i^1\left[
e^{\ii(B_i\phi_B+S_i\phi_S+Q_i\phi_Q)}
+ e^{-\ii(B_i\phi_B+S_i\phi_S+Q_i\phi_Q)}
\right]\right\}.
\label{pro1}
\end{eqnarray}
\end{widetext}
The direct numerical computation of the triple integral above is a formidable
task for $B \gtrsim 5$. The integral can be turned into a sum over many indices 
of modified Bessel functions $I_n(x)$, yet its evaluation is very time consuming
and becomes impractical with baryon number larger than $\sim 10$
\cite{cleymans2747}. Therefore, we have chosen another approach, which has been
applied in elementary collisions \cite{becattini269,becapass}. From (\ref{pro1})
one finds the integrand to be violently oscillating for large net quantum
numbers, 
so it is worth trying to eliminate analytically the source of strongest 
oscillation; in heavy ion reactions, 
the baryon number is always the largest of the set $B, S, Q$, so it is certainly
the most beneficial to eliminate the integration over $B$. In fact, this can be
done 
by taking advantage of a special feature of baryon number, that is no elementary
hadron exists with $|B_j|>1$. We first rewrite equation (\ref{pro1}) in the form
\begin{widetext}
\begin{eqnarray}
Z_{B,S,Q}(T,V) &=&  
Z_0 \frac{1}{(2\pi)^3}  \int_0^{2\pi}\dii\phi_B \; e^{-\ii B\phi_B}
\int_0^{2\pi}\dii\phi_S \; e^{-\ii S\phi_S}
\int_0^{2\pi}\dii\phi_Q \; e^{-\ii Q\phi_Q} \nonumber \\
&\times& \exp\left[\sum_{B,\overline{B}}z_i^1
e^{\ii(B_i\phi_B+S_i\phi_S+Q_i\phi_Q)}\right]
\exp\left[\sum_{M,\overline{M}}z_i^1
e^{\ii(S_i\phi_S+Q_i\phi_Q)}\right],\label{canpart}
\end{eqnarray}
\end{widetext}
where the two summations run over baryons and mesons, respectively. As $|B_i|=1$
for all of the baryons, we can write the baryon summation as:
\begin{eqnarray}
&&\sum_{B,\overline{B}}z_i^1 \,
e^{\ii(B_i\phi_B+S_i\phi_S+Q_i\phi_Q)} \ =  \\
&&e^{\ii\phi_B}\sum_B z_i^1 \, e^{\ii(S_i\phi_S+Q_i\phi_Q)} 
+ e^{-\ii\phi_B}\sum_B z_i^1 \, e^{-\ii(S_i\phi_S+Q_i\phi_Q)}, \nonumber
\end{eqnarray}
where baryonic and antibaryonic terms have been separated. Introducing the
notation 
$\sum_B z_i^1 \, e^{\ii(S_i\phi_S+Q_i\phi_Q)} = \omega $, the above equation can
be rewritten as:
\begin{eqnarray}
&& \sum_{B,\overline{B}}z_i^1 \,
e^{\ii(B_i\phi_B+S_i\phi_S+Q_i\phi_Q)}
\ =\ e^{\ii\phi_B}\omega + e^{-\ii\phi_B}\omega^* \nonumber \\
&&\ =\ e^{\ii(\phi_B+\arg \omega)}|\omega| 
+ e^{-\ii(\phi_B+\arg \omega)}|\omega|. 
\end{eqnarray}
Substituting this expression in Eq.~(\ref{canpart}) and changing the baryon 
variable according to $\phi_B \rightarrow \phi_B - \arg\omega$ allows us to
perform analytically the integration in $\phi_B$, which yields a modified Bessel
function I$_B$:
\begin{eqnarray}
Z_{B,S,Q}(T,V) &=& \frac{Z_0}{(2\pi)^2} \int_0^{2\pi}\dii\phi_S 
\int_0^{2\pi}\dii\phi_Q \label{partt}\\
&\times&\cos\left(S\phi_S+Q\phi_Q-B\arg \omega(\phi_S,\phi_Q) \right)
\nonumber \\ 
&\times&I_B(2|\omega(\phi_S,\phi_Q)|) \nonumber \\
&\times& \exp\left[2\sum_M z_i^1\cos(S_i\phi_S +
Q_i\phi_Q)\right]. \nonumber
\end{eqnarray}
Thus, we are left with a double integration only which can be then performed 
numerically with no major problem provided that, $B$ being very large, the
uniform asymptotic expansion of Bessel functions for large orders is used to 
compute I$_B$. 


It is worth noting, that the form (\ref{partt}) does not involve
a saddle-point approximation used in
\cite{bohr-mott,jennings440,bec-sollf}.
Actually, for very small systems it is not even applicable, because
the integrand (\ref{pro1}) is rather smoothly distributed
in the group parameter
space $(\phi_B,\phi_S,\phi_Q)$,
 so the basic assumption of the saddle-point method fails.


The mean particle numbers of primary hadrons (i.e. those directly emitted from
the hadronising source) $\langle N_i \rangle$ are obtained from the 
equation~(\ref{partt}) \cite{hagedorn541,becattini269,dasgupta1361}
by taking the derivative 
of the canonical partition 
function with respect to a fictitious fugacity $\lambda_i$:
\begin{eqnarray} 
\langle N_i \rangle &=& \left. \lambda_i \frac{\partial
\ln Z_{B,S,Q}(T,V)} {\partial \lambda_i} \right|_{\lambda_i =1} 
\nonumber \\
&=& \frac{Z_{B-B_i,S-S_i,Q-Q_i}(T,V)}{Z_{B,S,Q}(T,V)} z_i^1,
\label{Cmean}
\end{eqnarray}
where the quantity $Z_{B-B_i,S-S_i,Q-Q_i}/Z_{B,S,Q}$ is called {\em chemical
factor} 
\cite{becattini269}. In the large volume (thermodynamical) limit, this
expression
becomes the GC one, i.e. 
$\langle N_i \rangle = \lambda_B^{B_i}\lambda_S^{S_i}\lambda_Q^{Q_i}z_i^1$.

The canonical baryon chemical factor can be defined as that relevant to baryons
with vanishing strangeness and electric charge, e.g. neutrons: 
\begin{equation} \label{Cb}
C_B\ =\
\frac{Z_{B-1,S,Q}(T,V)}{Z_{B,S,Q}(T,V)}.
\end{equation}
so that in the GC limit one gets:
\begin{equation}
\lim_{V\rightarrow\infty}C_B\ =\ \lambda_B.
\end{equation}
Similarly, one can define chemical factors $C_S$ and $C_Q$ for strangeness and 
electric charge respectively. It must be emphasized that, unlike in the GC
framework 
where the total fugacity is actually a product of different 
charge fugacities powered to the number of charges carried by the hadron, the 
canonical chemical factors cannot be factorized and must be calculated, as
indicated 
in equation (\ref{Cmean}), as a ratio of partition functions for different sets
of quantum numbers. The relative differences between chemical factors and their
GC 
counterparts and quantities related to these differences are
defined, in this paper, 
as {\em canonical effects}.

In addition to the comparison between full canonical calculations and GC ones, 
it is very important to assess whether the enforcement of exact strangeness
conservation 
with GC treatment of baryon number and electric charge ({\em strangeness
canonical
ensemble}) is indeed sufficient to 
produce results 
close enough to full canonical ones. Here, we only quote the analytical form of 
strangeness-canonical partition function $Z_S$ derived using the method
introduced 
in \cite{cleymans137}:
\begin{eqnarray}
&&Z_S(T,V,\lambda_B,\lambda_Q) \ = \\ 
&&Z_0 \sum_{m=-\infty}^{\infty}I_m(2\zet_{\Xi^0})\lambda_B^m 
\sum_{n=-\infty}^{\infty}I_n(2\zet_{\Xi^-})\lambda_B^n\lambda_Q^{-n}
\nonumber \\
&&\times\sum_{l=-\infty}^{\infty}I_l(2\zet_{\Omega^-})\lambda_B^l\lambda_Q^{-l}
\nonumber \\
&&\times
\left(\sqrt{\frac{\en_1}{\en_{-1}}}\right)^{S+2m+2n+3l}
I_{S+2m+2n+3l}(2\sqrt{\en_1\en_{-1}}). \nonumber 
\end{eqnarray}
In the equation above, we denote the sum of one particle partition functions
carrying the same quantum numbers as $\Xi^0$ by $\zet_{\Xi^0}$, and similarly
for 
other kinds of multistrange hadrons, while $\en_j$ stands for the sum
$\sum_{S_i=j} 
\lambda_B^{B_i} \lambda_Q^{Q_i}z_i^1$. In this case, mean particle numbers read:
\begin{equation}
\langle N_i \rangle = \frac{Z_{S-S_i}(T,V,\lambda_B,\lambda_Q)}
{Z_S(T,V,\lambda_B,\lambda_Q)}\lambda_B^{B_i}
\lambda_Q^{Q_i}z^1_i.
\end{equation}

\section{Numerical results}\label{results}

First of all, we point out that we do not perform fits to experimental results 
in this paper. Indeed, as our main motivation is to test the statistical 
canonical formalism in the heavy ion collision regime, we only take some
suggestive 
values for freeze-out temperatures and baryon densities from our recent work on
the analysis of particle multiplicities in heavy ion systems in the laboratory 
momentum range from GSI SIS 1.7 $A$ GeV to CERN SPS 158 $A$ GeV \cite{features}
and
make comparisons between canonical and grand-canonical calculations for those
values. 

The peculiar feature of canonical formalism is a nonlinear volume behaviour of
mean particle numbers because of the dependence of chemical factors on the 
volume according to eqs.~(\ref{Cmean},\ref{pro1}). To show this effect, we first
fix the baryon density $n_B$ so that the dependence of the chemical factors on
the volume turns into a dependence on net baryon number. As the temperature is 
fixed too, then the grand-canonical baryon chemical potential can be calculated 
and the difference with the corresponding 
chemical factor can be studied. Furthermore, net strangeness $S$ and electric
charge $Q$ are fixed by the initial conditions of the collision (namely zero 
and $(Z/A) \times B$ respectively).

\begin{figure} 

\includegraphics[scale=0.34, angle=-90]{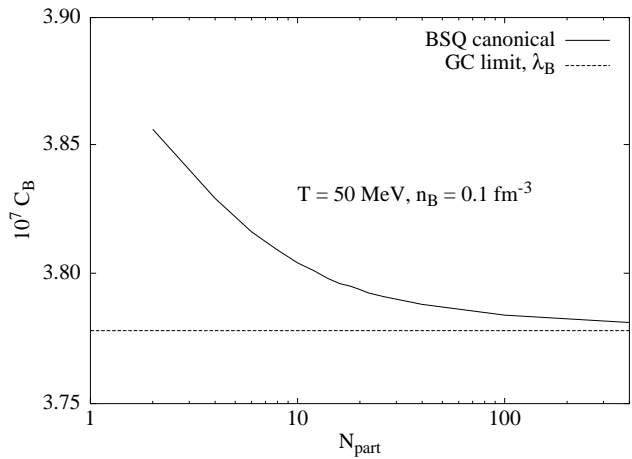}
\caption{Canonical baryon chemical factor as a function of
number of participants in the conditions relevant to GSI
energies. Thermodynamical limit is shown as a dashed line.}
\label{fig1}
\end{figure}

\begin{figure}
\includegraphics[scale=0.34, angle=-90]{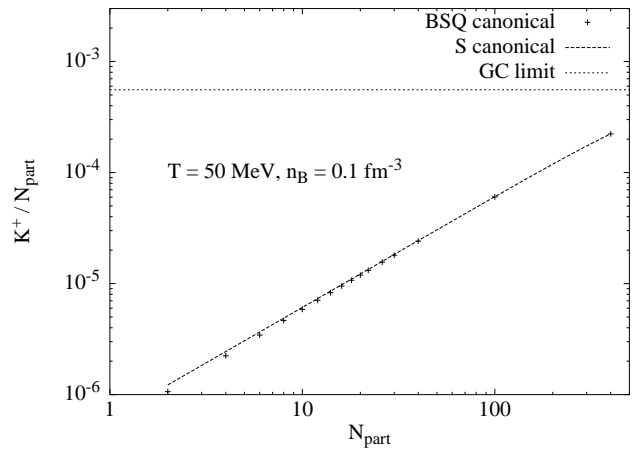}
\caption{Canonical enhancement of kaons as a function of number of participants 
in the conditions relevant to GSI energies.}
\label{fig2}
\end{figure}

The canonical effect is the most significant at comparatively low energies, 
such as 1.7 GeV per nucleon in SIS Au--Au collisions. In our recent analysis 
\cite{features} we find the best fit to experimental results with thermal
parameters 
$T\sim 50$ MeV and $\mu_B \sim 800$ MeV. These values yield, in the pointlike
hadron gas framework, a baryon density lower than 0.1 fm$^{-3}$, which is used 
as a reference value in order to study 
the canonical effect in heavy ion collisions as a function of the number of 
participants - i.e. net baryon number - in the collision. In figure \ref{fig1} 
the chemical factor $C_B$ is shown in comparison with its GC limit $\lambda_B$.
Under these circumstances, the canonical effect related to the baryon
chemical factor is found to decrease from 2.1\% at $B=2$ down to 0.7\% at
$B=10$, 
and to further decrease rather slowly towards the GC limit with increasing 
baryon number or volume. For central Au--Au reactions, the number of
participating 
nucleons is more than 300, so we can conclude that the canonical baryon chemical
effect is negligible. However, this is not the case for strangeness. In figure
\ref{fig2} the theoretical production ratio K$^+$ over number of participating 
nucleons is shown to point out the canonical strangeness effect. Note that we
have been able to compute the full chemical factor even for $B=400$.
The ratio increases very slowly towards the GC limit, thus strangeness must
always be 
handled canonically when analysing SIS results. The relative difference between 
full canonical and strangeness-canonical results is 13\% for $B=2$, decreasing
to 4\% for
$B=10$ and to 1\% for $B=40$. It must be pointed out that the above results are 
calculated using an isospin symmetric initial configuration whereas $Z/A\simeq 
0.4$ for gold nucleus. However, this variation essentially gives no change on
the
relative differences above, although the ratio K$^+/N_{\mbox{{\tiny part}}}$ 
decreases naturally due to the initial neutron excess.  

\begin{figure}
\includegraphics[scale=0.34, angle=-90]{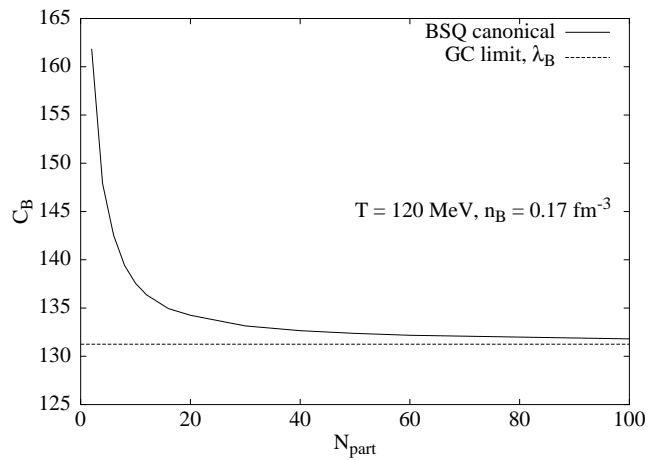}
\caption{Canonical baryon chemical factor as a function of the number of 
participants in the conditions relevant to AGS energies. Thermodynamical limit 
is shown as a dashed line.}
 \label{fig3}
\end{figure}

In Au--Au collisions at AGS energies (11.6 $A$ MeV momentum), the chemical 
freeze-out temperature is found to be around $T=120$ MeV and the baryon density 
close to the normal nuclear 
density \cite{features}. Using these values, we show the canonical baryon
chemical 
factor in comparison with the GC limit in figure~\ref{fig3} keeping a symmetric
isospin initial condition.  
Relative difference between $C_B$ and $\lambda_B$ decreases rather quickly 
from 23\% at $B=2$ to 1\% at $B=40$. A more realistic comparison of results 
in Au--Au with pp collisions requires different initial isospin configuration 
to be into account. By using temperature and baryon density quoted 
above, $C_B$ at $B=Q=2$ turns out to be 81.9 and $\lambda_B = 118$ with
$Q/B=0.4$.

\begin{figure}
\includegraphics[scale=0.34, angle=-90]{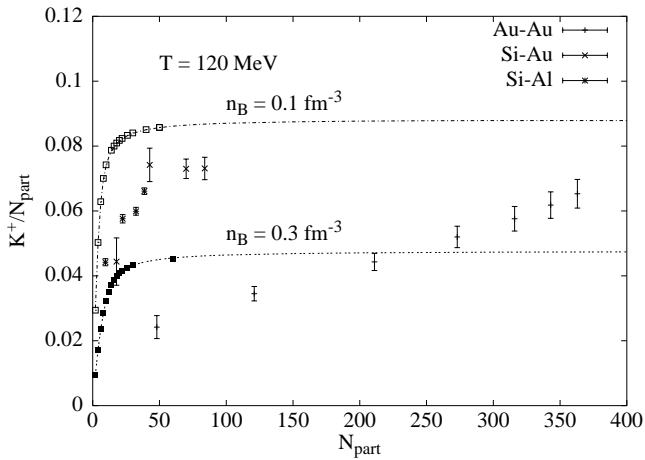}
\caption{Theoretical K$^+/N_{\mbox{{\tiny part}}}$ curves at fixed temperature 
for two different baryon densities shown along with AGS experimental results 
\cite{ahle}. Curves are strangeness canonical while squares are full canonical
results}
\label{fig4}
\end{figure}

\begin{figure}
\includegraphics[scale=0.34, angle=-90]{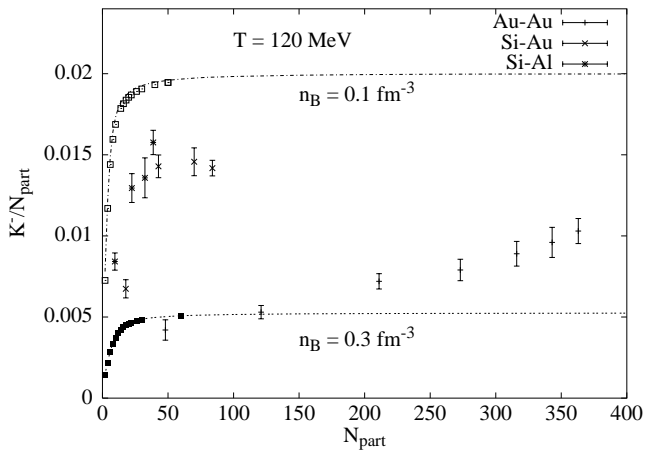}
\caption{Theoretical K$^-/N_{\mbox{{\tiny part}}}$ curves at fixed temperature 
for two different baryon densities shown along with AGS experimental results 
\cite{ahle}. The curves are strangeness canonical while squares are full
canonical results}
\label{fig5}
\end{figure}

Recent experimental results on AGS K$/N_{\mbox{{\tiny part}}}$ ratios with 
increasing centrality \cite{ahle} serve as a basis for a further study of
strangeness production behaviour at $T=120$ MeV. In figure \ref{fig4} we plot
the 
calculated K$^+/N_{\mbox{{\tiny part}}}$ with results obtained in peripheral to 
central Au--Au, Si--Au and Si--Al collisions.
It can be seen that all the way up from $B=2$ to $B=60$ the full canonical and
strangeness-canonical results are essentially the same. All central reaction 
results lie on the region where the canonical effects are negligible and the GC 
formalism applies.
Strangeness enhancement in Au--Au system does not definitely look like being
of canonical origin, whereas Si--Au and Si--Al follow roughly the curve at the 
normal nuclear density (not shown in the figure). The above argument also
applies for the K$^-/N_{\mbox{{\tiny part}}}$ enhancement pattern shown in
figure~\ref{fig5}.

Whereas the kaon enhancement with increasing centrality seems not to be
compatible 
with a purely canonical effect, inelastic pp and {\em central} p--A and A--A
systems 
considered in ref.~\cite{cleymans2747} show a different behaviour with
increasing 
volume. Therein, it was found that K$/\pi$ ratios clearly follow the canonical
curves. 

\begin{figure}
\includegraphics[scale=0.34, angle=-90]{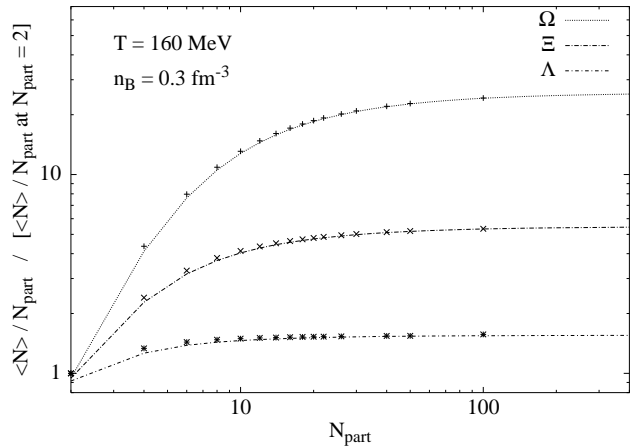}
\caption{Strange baryon enhancement in the conditions relevant to the SPS
energies. 
Hadron multiplicities are normalized to full canonical
results at $B=2$. Curves are strangeness 
canonical while crosses are full canonical results.} 
\label{fig6}
\end{figure}

The best fit temperature for the multiplicities measured by NA49 experiment at
SPS 
in central Pb--Pb reactions at 158 $A$ GeV is found to be $T\sim160$ MeV
\cite{features}.
In figure \ref{fig6} we show the strange baryon enhancement by using a baryon
density 
0.3 fm$^{-3}$. Although the baryon density in our analysis \cite{features} was
found
to be $n_B \sim 0.2$ fm$^{-3}$ this larger value is used in order to probe the 
largest reasonable canonical effect. 
All results in figure \ref{fig6} are normalized to the {\em full canonical} 
    baryon multiplicities per participant at the point
$N_{\mbox{{\tiny part}}} = 2$. This choice reveals the slight
    difference between the results obtained using the $S$-canonical
    approximation and the $\bsq$-canonical calculation,
 and forces
both canonical methods to reach for the same GC limit.
An interesting feature here is the fact, that the $S$-canonical
method leads to an overestimation of the canonical effect.  
Corresponding curves 
have also been calculated in
ref.~\cite{redlich413} within the strangeness canonical ensemble with some
numerical approximations not used here. This 
enhancement picture has been calculated again with the assumption of an initial 
symmetric isospin configuration. 
Changing the freeze-out conditions to $T=160$ MeV and $n_B = 0.17$ fm$^-3$,
the total relative enhancement of $\langle N \rangle / N_{\mbox{{\tiny part}}}$
from
canonical $B=Q=2$ to GC $Q/B = 0.4$ raises to 30\% for $\Lambda$'s, 241\% for 
$\Xi$'s and 918\% for $\Omega$'s.  

\begin{figure}
\includegraphics[scale=0.34, angle=-90]{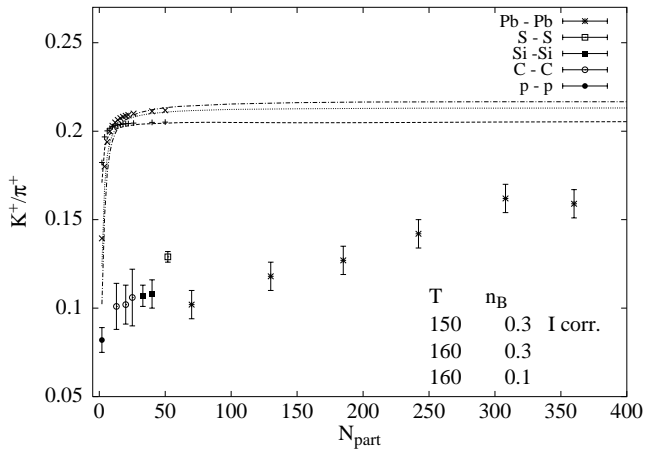}
\caption{Theoretical K$^+/\pi^+$ curves at fixed temperature compared with SPS 
experimental results \cite{blume}. Curves are strangeness canonical while
crosses are full canonical results. Curves from top to bottom are indicated in
the
lower right corner. Crosses correspond to the curves with $T=160$ MeV.}
\label{fig7}
\end{figure}

SPS NA49 collaboration has also measured the participant dependence of
K$^+/\pi^+$ 
ratio in pp, C--C, Si--Si, S--S and Pb--Pb reactions \cite{blume}. In
fig.~\ref{fig7} 
we show that the roughly linear enhancement pattern is far away from the purely 
canonical shape. 
In the same figure, one can see the weak dependence of the K$^+/\pi^+$ ratio on
thermal parameters. Indeed, theoretical curves always lie well above the
measured points, which is a strong indication that strangeness is not in
complete
chemical equilibrium. Usually, this lack of equilibrium 
is taken into account by introducing a parameter $\gamma_S$ (see discussion in
section~\ref{gamma}), 
which, in the Boltzmann approximation, appears as a linear coefficient in front 
of the K$/\pi$ ratios. Taking our best fit value $\gamma_S \sim 0.8$
\cite{features} 
brings the theoretical curves asymptotically to the experimental central Pb--Pb 
reaction points. Curves with $T=160$ MeV (lowest two) assume initial isospin
symmetry, 
whilst the $S$ canonical curve with $T=150$ MeV is corrected for the condition 
$Q/B = 0.4$. It is worth quoting our result for pp to central Pb--Pb K$^+/\pi^+
= R$ enhancement using $T=160$ MeV and $n_B = 0.17$ fm$^{-3}$ taking into
account
the initial isospin: $R_{\mbox{{\tiny Pb--Pb}}}/R_{\mbox{{\tiny pp}}} = 1.47$
with 
$\gamma_S = 1$. Experimentally, this ratio is about 2 as it can be inferred by
looking at the points in fig.~\ref{fig7}.

\section{Discussion on the relative chemical equilibrium}
\label{gamma}

In statistical model analysis of heavy ion reactions it is found that the 
assumption of full chemical equilibrium of strangeness is mostly not satisfied 
(\cite{features} and this work). Thus, a concept of relative chemical
equilibrium was
inroduced \cite{koch167} and has been applied in many subsequent works. The
relative 
chemical equilibrium means that the strange quark production is not fully 
equilibrated, but the distribution of strange quarks among different hadron
species 
occurs according to statistical equilibrium ({\em statistical coalescence}).

In order to parametrize the relative chemical equilibrium, another constraint in
the derivation of the relevant statistical operator is needed. In the GC
formalism, this constraint is the conservation of the average number of strange
and 
antistrange quarks in addition to usual constraints for energy and charge
conservation. 
Note that this constraint is not owing to any natural symmetry of strong
interactions
like baryon number, electric charge and strangeness conservation. Hence, as the
number
of produced strange quarks to be distributed among final hadrons is unknown, it 
always leads to the introduction of another free parameter to be determined
{\em a posteriori.}

Using the aforementioned constraints, the grand canonical (GC) partition
function 
can be written as:
\begin{equation}\label{ZGC}
Z_{\rm{GC}}(T,V,\mu,\mu_{n_s}) =
\Tr\left[e^{-\beta(H-\mu N-\mu_{n_s}n_s)}\right],
\end{equation}

where the chemical potentials for all the relevant charges have been grouped 
under the symbol $\mu$. The fugacity factor for the constrained number of
strange 
quarks $e^{\beta\mu_{n_s}n_s}$ is usually called $\gamma_S$
\cite{koch167,rafelski333}.

In the full canonical picture, where charges are not allowed to fluctuate from 
sample to sample, the constraint of the average number of strange and
antistrange quarks can be turned into the enforcement of a fixed {\em exact} 
number of $s,\bar{s}$ quarks. Corresponding $U(1)_{n_s}$ symmetry yields another
integration variable and another phase factor, 
$
\exp(\ii n_s \phi_{n_s}),
$
into the projection integral (\ref{pro1}) \cite{becapass}. Following the
procedure 
described in section~\ref{model} the average number of hadrons $h_i$ can be 
written as:
\begin{equation}\label{cangs}
<h_i> = \frac{Z_{B-B_i,S-S_i,Q-Q_i,n_s-(n_s)_i}}
{Z_{B,S,Q,n_s}} z^1_i.
\end{equation}
In the grand canonical limit, the corresponding number is
\begin{equation}\label{gcgs}
\lim_{V\rightarrow\infty}<h_i> = \lambda_B^{B_i}\lambda_S^{S_i}
\lambda_Q^{Q_i}\gamma_S^{n_{si}}z^1_i.
\end{equation} 
As an illustrative example, take the average number of 
$\phi$ mesons. It does not carry any relevant charge, but
the valence quark content is $s\bar{s}$. In the GC limit
$<\phi> = \gamma_S^2 z^1_\phi$. This leads us to the identification
\begin{equation}
\lim_{V\rightarrow\infty}\frac{Z_{B,S,Q,n_s-2}}
{Z_{B,S,Q,n_s}} = \gamma_S^2.
\end{equation}

There is one major caveat in those formulae and the use of $\gamma_S$.
Indeed, the canonical formalism for the {\em net conserved charges} in the
system is to be applied in any reaction where they are known from the initial
state; if the system is large enough, the fluctuations allowed in GC ensemble
give negligible deviation from canonical results and one is allowed to use
formulae such as eq.~(\ref{gcgs}). However, the case of the number of strange
quarks $N_s$ is instrinsically different. In fact, this number is not known
from the initial state and may undergo large dynamical fluctuations from 
event to event, well beyond those (small) predicted by the GC formalism and 
its related fugacity $\gamma_S$. If this was the case, $\gamma_S$ in 
eq.~(\ref{gcgs}) must be
understood as a sort of {\em average} fugacity rather than a proper fugacity.
On the other hand, when the expected mean number of $s\bar{s}$ pairs is small,
the canonical formalism should be used (i.e. eq.~(\ref{cangs}))
but the probability distribution of generating a given number of $s\bar{s}$
pairs should be known in advance. As a reasonable {\em ansatz}, one can assume
those pairs are independently produced so that their distribution is a
Poissonian
\cite{becapass}.

\section{Conclusions}

We have presented a method to calculate particle densities in the canonical 
framework of the statistical model, which can be effectively used in 
heavy ion reactions at very large values of baryon number. This method has 
allowed us to study the applicability of different approximations in chemical 
analysis, namely the grand-canonical and the strangeness-canonical ensembles.
It is found that the full $\bsq$ canonical formalism is only needed for 
elementary and very small nuclear systems up to baryon number $\sim 10$. For 
larger systems, the baryon number and the electric charge can be safely
handled in grand-canonical manner. For central Au--Au and Pb--Pb reactions, 
exact strangeness conservation is needed at GSI SIS energies 
($p_{\mbox{\tiny{lab}}} \leq 2$ GeV A), while at AGS and SPS even strangeness
can be handled grand-canonically. Certainly, this also applies to much higher 
energy RHIC and LHC collisions.

The observed strangeness enhancement from peripheral to central nuclear 
collisions \cite{ahle,blume} is generally not reproduced by the canonical curves
with fixed temperature and baryon density whilst the AGS p--p to p--A to A--A 
{\em central} reaction measurements are found to follow the canonical results 
\cite{cleymans2747}. This is an indication of a possible change in reaction 
systematics going from peripheral to central collisions.
The strange hadron enhancement in SPS from p--p to Pb--Pb reactions is only 
qualitatively reproduced by canonical model.

Whilst $\gamma_S$ was not used in the derivation of our numerical results, 
the weak dependence of K$/\pi$ ratio on the thermal parameters 
(shown in fig.~\ref{fig7}) serves as a further compelling piece of evidence
in favour of incomplete strangeness chemical eqilibration in CERN SPS heavy 
ion reactions, confirming previous findings \cite{bec-sollf,features}.\\

\acknowledgments
We would like to acknowledge stimulating discussions
with Esko Suhonen, Jean Cleymans, Krzysztof Redlich, Ulrich Heinz and
L\'aszl\'o Csernai. 
One of us (A.K.) 
acknowledges the support of the Bergen Computational
Physics Laboratory in the framework of the European Community - Access
to Research Infrastructure action of the
Improving Human Potential Programme. 



\end{document}